\documentclass[%
    superscriptaddress,
    reprint,
    showpacs,preprintnumbers,
    nofootinbib,
    amsmath,amssymb,
    aps,
    prl, 
    floatfix,
    colorlinks,       
    citecolor=blue,   
    urlcolor=blue,    
    linkcolor=blue    
]{revtex4-2}

\usepackage[table,xcdraw]{xcolor}

\usepackage{float}

\usepackage{graphicx}
\usepackage{dcolumn}
\usepackage{bm}

\usepackage{afterpage}
\usepackage{tikz}
\usepackage{textgreek}

\usepackage{caption}
\usepackage{orcidlink}
\usepackage{todonotes}

\usepackage{xspace}

\usepackage{lineno}
\usepackage{soul}

\begin{document}

\title{First measurement of $\pi^+$--Ar and $p$--Ar total inelastic cross sections in the sub-GeV energy regime with ProtoDUNE-SP data}

%

\newcommand{\Albanysuny}{University of Albany, SUNY, Albany, NY 12222, USA}
\newcommand{\Almaty}{Institute of Nuclear Physics at Almaty, Almaty 050032, Kazakhstan
}
\newcommand{\Amsterdam}{University of Amsterdam, NL-1098 XG Amsterdam, The Netherlands}
\newcommand{\Antalya}{Antalya Bilim University, 07190 D\"o{\c s}emealtı/Antalya, Turkey}
\newcommand{\Antananarivo}{University of Antananarivo, Antananarivo 101, Madagascar}
\newcommand{\Antioquia}{University of Antioquia, Medell\'in, Colombia}
\newcommand{\AntonioNarino}{Universidad Antonio Nari\~no, Bogot\'a, Colombia}
\newcommand{\Argonne}{Argonne National Laboratory, Argonne, IL 60439, USA}
\newcommand{\Arizona}{University of Arizona, Tucson, AZ 85721, USA}
\newcommand{\Asuncion}{Universidad Nacional de Asunci\'on, San Lorenzo, Paraguay}
\newcommand{\Athens}{University of Athens, Zografou GR 157 84, Greece}
\newcommand{\Atlantico}{Universidad del Atl\'antico, Barranquilla, Atl\'antico, Colombia}
\newcommand{\Augustana}{Augustana University, Sioux Falls, SD 57197, USA}
\newcommand{\Bern}{University of Bern, CH-3012 Bern, Switzerland}
\newcommand{\Beykent}{Beykent University, Istanbul, Turkey}
\newcommand{\Birmingham}{University of Birmingham, Birmingham B15 2TT, United Kingdom}
\newcommand{\BolognaUniversity}{Universit\`a di Bologna, 40127 Bologna, Italy}
\newcommand{\Boston}{Boston University, Boston, MA 02215, USA}
\newcommand{\Bristol}{University of Bristol, Bristol BS8 1TL, United Kingdom}
\newcommand{\Brookhaven}{Brookhaven National Laboratory, Upton, NY 11973, USA}
\newcommand{\Bucharest}{University of Bucharest, Bucharest, Romania}
\newcommand{\CalBerkeley}{University of California Berkeley, Berkeley, CA 94720, USA}
\newcommand{\CalDavis}{University of California Davis, Davis, CA 95616, USA}
\newcommand{\CalIrvine}{University of California Irvine, Irvine, CA 92697, USA}
\newcommand{\CalLosangeles}{University of California Los Angeles, Los Angeles, CA 90095, USA}
\newcommand{\CalRiverside}{University of California Riverside, Riverside CA 92521, USA}
\newcommand{\CalSantabarbara}{University of California Santa Barbara, Santa Barbara, CA 93106, USA}
\newcommand{\Caltech}{California Institute of Technology, Pasadena, CA 91125, USA}
\newcommand{\Cambridge}{University of Cambridge, Cambridge CB3 0HE, United Kingdom}
\newcommand{\Campinas}{Universidade Estadual de Campinas, Campinas - SP, 13083-970, Brazil}
\newcommand{\CataniaUniversitadi}{Universit\`a di Catania, 2 - 95131 Catania, Italy}
\newcommand{\Catolica}{Universidad Cat\'olica del Norte, Antofagasta, Chile}
\newcommand{\CBPF}{Centro Brasileiro de Pesquisas F\'isicas, Rio de Janeiro, RJ 22290-180, Brazil}
\newcommand{\CEASaclay}{IRFU, CEA, Universit\'e Paris-Saclay, F-91191 Gif-sur-Yvette, France}
\newcommand{\CERN}{CERN, The European Organization for Nuclear Research, 1211 Meyrin, Switzerland}
\newcommand{\Charles}{Institute of Particle and Nuclear Physics of the Faculty of Mathematics and Physics of the Charles University, 180 00 Prague 8, Czech Republic }
\newcommand{\Chicago}{University of Chicago, Chicago, IL 60637, USA}
\newcommand{\ChungAng}{Chung-Ang University, Seoul 06974, South Korea}
\newcommand{\CIEMAT}{CIEMAT, Centro de Investigaciones Energ\'eticas, Medioambientales y Tecnol\'ogicas, E-28040 Madrid, Spain}
\newcommand{\Cincinnati}{University of Cincinnati, Cincinnati, OH 45221, USA}
\newcommand{\Cinvestav}{Centro de Investigaci\'on y de Estudios Avanzados del Instituto Polit\'ecnico Nacional (Cinvestav), Mexico City, Mexico}
\newcommand{\Colima}{Universidad de Colima, Colima, Mexico}
\newcommand{\ColoradoBoulder}{University of Colorado Boulder, Boulder, CO 80309, USA}
\newcommand{\ColoradoState}{Colorado State University, Fort Collins, CO 80523, USA}
\newcommand{\Columbia}{Columbia University, New York, NY 10027, USA}
\newcommand{\conida}{Comisi\'on Nacional de Investigaci\'on y Desarrollo Aeroespacial, Lima, Peru}
\newcommand{\Cti}{Centro de Tecnologia da Informacao Renato Archer, Amarais - Campinas, SP - CEP 13069-901}
\newcommand{\CUSB}{Central University of South Bihar, Gaya, 824236, India
}
\newcommand{\CzechAcademyofSciences}{Institute of Physics, Czech Academy of Sciences, 182 00 Prague 8, Czech Republic}
\newcommand{\CzechTechnical}{Czech Technical University, 115 19 Prague 1, Czech Republic}
\newcommand{\DannecyleVieux}{Laboratoire d'Annecy de Physique des Particules, Universit\'e Savoie Mont Blanc, CNRS, LAPP-IN2P3, 74000 Annecy, France}
\newcommand{\Daresbury}{Daresbury Laboratory, Cheshire WA4 4AD, United Kingdom}
\newcommand{\Dordt}{Dordt University, Sioux Center, IA 51250, USA}
\newcommand{\Drexel}{Drexel University, Philadelphia, PA 19104, USA}
\newcommand{\Duke}{Duke University, Durham, NC 27708, USA}
\newcommand{\Durham}{Durham University, Durham DH1 3LE, United Kingdom}
\newcommand{\Edinburgh}{University of Edinburgh, Edinburgh EH8 9YL, United Kingdom}
\newcommand{\EIA}{Universidad EIA, Envigado, Antioquia, Colombia}
\newcommand{\Eotvos}{E\"otv\"os Lor\'and University, 1053 Budapest, Hungary}
\newcommand{\erciyes}{Erciyes University, Kayseri, Turkey}
\newcommand{\FCULport}{Faculdade de Ci\^encias da Universidade de Lisboa - FCUL, 1749-016 Lisboa, Portugal}
\newcommand{\FederaldeAlfenas}{Universidade Federal de Alfenas, Po{\c c}os de Caldas - MG, 37715-400, Brazil}
\newcommand{\FederaldeGoias}{Universidade Federal de Goias, Goiania, GO 74690-900, Brazil}
\newcommand{\FederaldoABC}{Universidade Federal do ABC, Santo Andr\'e - SP, 09210-580, Brazil}
\newcommand{\FederaldoRio}{Universidade Federal do Rio de Janeiro, Rio de Janeiro - RJ, 21941-901, Brazil}
\newcommand{\Fermi}{Fermi National Accelerator Laboratory, Batavia, IL 60510, USA}
\newcommand{\Ferrarauniv}{University of Ferrara, Ferrara, Italy}
\newcommand{\Florida}{University of Florida, Gainesville, FL 32611-8440, USA}
\newcommand{\Floridastate}{Florida State University, Tallahassee, FL, 32306 USA}
\newcommand{\Fluminense}{Fluminense Federal University, 9 Icara\'i Niter\'oi - RJ, 24220-900, Brazil }
\newcommand{\Genova}{Universit\`a degli Studi di Genova, Genova, Italy}
\newcommand{\Georgian}{Georgian Technical University, Tbilisi, Georgia}
\newcommand{\Granada}{University of Granada \& CAFPE, 18002 Granada, Spain}
\newcommand{\GranSasso}{Gran Sasso Science Institute, L'Aquila, Italy}
\newcommand{\GranSassoLab}{Laboratori Nazionali del Gran Sasso, L'Aquila AQ, Italy}
\newcommand{\Grenoble}{University Grenoble Alpes, CNRS, Grenoble INP, LPSC-IN2P3, 38000 Grenoble, France}
\newcommand{\Guanajuato}{Universidad de Guanajuato, Guanajuato, C.P. 37000, Mexico}
\newcommand{\Harish}{Harish-Chandra Research Institute, Jhunsi, Allahabad 211 019, India}
\newcommand{\Hawaii}{University of Hawaii, Honolulu, HI 96822, USA}
\newcommand{\hkust}{Hong Kong University of Science and Technology, Kowloon, Hong Kong, China}
\newcommand{\Houston}{University of Houston, Houston, TX 77204, USA}
\newcommand{\Hyderabad}{University of  Hyderabad, Gachibowli, Hyderabad - 500 046, India}
\newcommand{\Idaho}{Idaho State University, Pocatello, ID 83209, USA}
\newcommand{\IFIC}{Instituto de F\'isica Corpuscular, CSIC and Universitat de Val\`encia, 46980 Paterna, Valencia, Spain}
\newcommand{\IGFAE}{Instituto Galego de F\'isica de Altas Enerx\'ias, University of Santiago de Compostela, Santiago de Compostela, 15782, Spain}
\newcommand{\ihep}{Institute of High Energy Physics, Chinese Academy of Sciences, Beijing, China}
\newcommand{\Iitk}{Indian Institute of Technology Kanpur, Uttar Pradesh 208016, India}
\newcommand{\Illinoisinstitute}{Illinois Institute of Technology, Chicago, IL 60616, USA}
\newcommand{\Imperial}{Imperial College of Science, Technology and Medicine, London SW7 2BZ, United Kingdom}
\newcommand{\IndGuwahati}{Indian Institute of Technology Guwahati, Guwahati, 781 039, India}
\newcommand{\IndHyderabad}{Indian Institute of Technology Hyderabad, Hyderabad, 502285, India}
\newcommand{\Indiana}{Indiana University, Bloomington, IN 47405, USA}
\newcommand{\INFNBologna}{Istituto Nazionale di Fisica Nucleare Sezione di Bologna, 40127 Bologna BO, Italy}
\newcommand{\INFNCatania}{Istituto Nazionale di Fisica Nucleare Sezione di Catania, I-95123 Catania, Italy}
\newcommand{\INFNFerrara}{Istituto Nazionale di Fisica Nucleare Sezione di Ferrara, I-44122 Ferrara, Italy}
\newcommand{\INFNFrascati}{Istituto Nazionale di Fisica Nucleare Laboratori Nazionali di Frascati, Frascati, Roma, Italy}
\newcommand{\INFNGenova}{Istituto Nazionale di Fisica Nucleare Sezione di Genova, 16146 Genova GE, Italy}
\newcommand{\INFNLecce}{Istituto Nazionale di Fisica Nucleare Sezione di Lecce, 73100 - Lecce, Italy}
\newcommand{\INFNMilanBicocca}{Istituto Nazionale di Fisica Nucleare Sezione di Milano Bicocca, 3 - I-20126 Milano, Italy}
\newcommand{\INFNMilano}{Istituto Nazionale di Fisica Nucleare Sezione di Milano, 20133 Milano, Italy}
\newcommand{\INFNNapoli}{Istituto Nazionale di Fisica Nucleare Sezione di Napoli, I-80126 Napoli, Italy}
\newcommand{\INFNPadova}{Istituto Nazionale di Fisica Nucleare Sezione di Padova, 35131 Padova, Italy}
\newcommand{\INFNPavia}{Istituto Nazionale di Fisica Nucleare Sezione di Pavia,  I-27100 Pavia, Italy}
\newcommand{\INFNPisa}{Istituto Nazionale di Fisica Nucleare Laboratori Nazionali di Pisa, Pisa PI, Italy}
\newcommand{\INFNRoma}{Istituto Nazionale di Fisica Nucleare Sezione di Roma, 00185 Roma RM, Italy}
\newcommand{\INFNRomavergata}{Istituto Nazionale di Fisica Nucleare Roma Tor Vergata , 00133 Roma RM, Italy}
\newcommand{\INFNSud}{Istituto Nazionale di Fisica Nucleare Laboratori Nazionali del Sud, 95123 Catania, Italy}
\newcommand{\Infntorino}{Istituto Nazionale di Fisica Nucleare, Sezione di Torino, Turin, Italy}
\newcommand{\Ingenieria}{Universidad Nacional de Ingenier\'ia, Lima 25, Per\'u}
\newcommand{\Insubria }{University of Insubria, Via Ravasi, 2, 21100 Varese VA, Italy}
\newcommand{\Iowa}{University of Iowa, Iowa City, IA 52242, USA}
\newcommand{\IowaState}{Iowa State University, Ames, Iowa 50011, USA}
\newcommand{\IPLyon}{Institut de Physique des 2 Infinis de Lyon, 69622 Villeurbanne, France}
\newcommand{\IPM}{Institute for Research in Fundamental Sciences, Tehran, Iran}
\newcommand{\IRLPPC}{Particle Physics and Cosmology International Research Laboratory	, Chicago IL,  60637 USA}
\newcommand{\ISTlisboa}{Instituto Superior T\'ecnico - IST, Universidade de Lisboa, 1049-001 Lisboa, Portugal}
\newcommand{\Ita}{Instituto Tecnol\'ogico de Aeron\'autica, Sao Jose dos Campos, Brazil}
\newcommand{\Iwate}{Iwate University, Morioka, Iwate 020-8551, Japan}
\newcommand{\Jacksonstate}{Jackson State University, Jackson, MS 39217, USA}
\newcommand{\Jawaharlal}{Jawaharlal Nehru University, New Delhi 110067, India}
\newcommand{\Jeonbuk}{Jeonbuk National University, Jeonrabuk-do 54896, South Korea}
\newcommand{\Jyvaskyla}{Jyv\"askyl\"a University, FI-40014 Jyv\"askyl\"a, Finland}
\newcommand{\Kansasstate}{Kansas State University, Manhattan, KS 66506, USA}
\newcommand{\Kavli}{Kavli Institute for the Physics and Mathematics of the Universe, Kashiwa, Chiba 277-8583, Japan}
\newcommand{\KEK}{High Energy Accelerator Research Organization (KEK), Ibaraki, 305-0801, Japan}
\newcommand{\KISTI}{Korea Institute of Science and Technology Information, Daejeon, 34141, South Korea}
\newcommand{\Kyiv}{Taras Shevchenko National University of Kyiv, 01601 Kyiv, Ukraine}
\newcommand{\Lancaster}{Lancaster University, Lancaster LA1 4YB, United Kingdom}
\newcommand{\LawrenceBerkeley}{Lawrence Berkeley National Laboratory, Berkeley, CA 94720, USA}
\newcommand{\LIP}{Laborat\'orio de Instrumenta{\c c}\~ao e F\'isica Experimental de Part\'iculas, 1649-003 Lisboa and 3004-516 Coimbra, Portugal}
\newcommand{\Liverpool}{University of Liverpool, L69 7ZE, Liverpool, United Kingdom}
\newcommand{\LosAlmos}{Los Alamos National Laboratory, Los Alamos, NM 87545, USA}
\newcommand{\Louisanastate}{Louisiana State University, Baton Rouge, LA 70803, USA}
\newcommand{\LpBordeaux}{Laboratoire de Physique des Deux Infinis Bordeaux - IN2P3, F-33175 Gradignan, Bordeaux, France, }
\newcommand{\Lucknow}{University of Lucknow, Uttar Pradesh 226007, India}
\newcommand{\Mainz}{Johannes Gutenberg-Universit\"at Mainz, 55122 Mainz, Germany}
\newcommand{\Manchester}{University of Manchester, Manchester M13 9PL, United Kingdom}
\newcommand{\Massinsttech}{Massachusetts Institute of Technology, Cambridge, MA 02139, USA}
\newcommand{\Medellin}{University of Medell\'in, Medell\'in, 050026 Colombia }
\newcommand{\Michigan}{University of Michigan, Ann Arbor, MI 48109, USA}
\newcommand{\Michiganstate}{Michigan State University, East Lansing, MI 48824, USA}
\newcommand{\MilanoBicocca}{Universit\`a di Milano Bicocca , 20126 Milano, Italy}
\newcommand{\MilanoUniv}{Universit\`a degli Studi di Milano, I-20133 Milano, Italy}
\newcommand{\Minnduluth}{University of Minnesota Duluth, Duluth, MN 55812, USA}
\newcommand{\Minntwin}{University of Minnesota Twin Cities, Minneapolis, MN 55455, USA}
\newcommand{\Mississippi}{University of Mississippi, University, MS 38677 USA}
\newcommand{\napoli}{Universit\`a degli Studi di Napoli Federico II , 80138 Napoli NA, Italy}
\newcommand{\Nikhef}{Nikhef National Institute of Subatomic Physics, 1098 XG Amsterdam, Netherlands}
\newcommand{\Niser}{National Institute of Science Education and Research (NISER), Odisha 752050, India}
\newcommand{\Northdakota}{University of North Dakota, Grand Forks, ND 58202-8357, USA}
\newcommand{\Northernillinois}{Northern Illinois University, DeKalb, IL 60115, USA}
\newcommand{\Northwestern}{Northwestern University, Evanston, Il 60208, USA}
\newcommand{\NotreDame}{University of Notre Dame, Notre Dame, IN 46556, USA}
\newcommand{\NoviSad}{University of Novi Sad, 21102 Novi Sad, Serbia}
\newcommand{\Ohiostate}{Ohio State University, Columbus, OH 43210, USA}
\newcommand{\OregonState}{Oregon State University, Corvallis, OR 97331, USA}
\newcommand{\Oxford}{University of Oxford, Oxford, OX1 3RH, United Kingdom}
\newcommand{\PacificNorthwest}{Pacific Northwest National Laboratory, Richland, WA 99352, USA}
\newcommand{\Padova}{Universt\`a degli Studi di Padova, I-35131 Padova, Italy}
\newcommand{\Panjab}{Panjab University, Chandigarh, 160014, India}
\newcommand{\Parissaclay}{Universit\'e Paris-Saclay, CNRS/IN2P3, IJCLab, 91405 Orsay, France}
\newcommand{\Parisuniversite}{Universit\'e Paris Cit\'e, CNRS, Astroparticule et Cosmologie, Paris, France}
\newcommand{\Parma}{University of Parma,  43121 Parma PR, Italy}
\newcommand{\Pavia}{Universit\`a degli Studi di Pavia, 27100 Pavia PV, Italy}
\newcommand{\Penn}{University of Pennsylvania, Philadelphia, PA 19104, USA}
\newcommand{\PennState}{Pennsylvania State University, University Park, PA 16802, USA}
\newcommand{\PhysicalResearchLaboratory}{Physical Research Laboratory, Ahmedabad 380 009, India}
\newcommand{\Pisa}{Universit\`a di Pisa, I-56127 Pisa, Italy}
\newcommand{\Pitt}{University of Pittsburgh, Pittsburgh, PA 15260, USA}
\newcommand{\Pontificia}{Pontificia Universidad Cat\'olica del Per\'u, Lima, Per\'u}
\newcommand{\PuertoRico}{University of Puerto Rico, Mayaguez 00681, Puerto Rico, USA}
\newcommand{\Punjab}{Punjab Agricultural University, Ludhiana 141004, India}
\newcommand{\QMUL}{Queen Mary University of London, London E1 4NS, United Kingdom
}
\newcommand{\Radboud}{Radboud University, NL-6525 AJ Nijmegen, Netherlands}
\newcommand{\Rice}{Rice University, Houston, TX 77005}
\newcommand{\Rochester}{University of Rochester, Rochester, NY 14627, USA}
\newcommand{\Royalholloway}{Royal Holloway College London, London, TW20 0EX, United Kingdom}
\newcommand{\Rutgers}{Rutgers University, Piscataway, NJ, 08854, USA}
\newcommand{\Rutherford}{STFC Rutherford Appleton Laboratory, Didcot OX11 0QX, United Kingdom}
\newcommand{\Salento}{Universit\`a del Salento, 73100 Lecce, Italy}
\newcommand{\santamarta}{Universidad del Magdalena, Santa Marta - Colombia}
\newcommand{\Sapienza}{Sapienza University of Rome, 00185 Roma RM, Italy}
\newcommand{\SergioArboleda}{Universidad Sergio Arboleda, 11022 Bogot\'a, Colombia}
\newcommand{\Sheffield}{University of Sheffield, Sheffield S3 7RH, United Kingdom}
\newcommand{\SLAC}{SLAC National Accelerator Laboratory, Menlo Park, CA 94025, USA}
\newcommand{\Southcarolina}{University of South Carolina, Columbia, SC 29208, USA}
\newcommand{\SouthDakotaSchool}{South Dakota School of Mines and Technology, Rapid City, SD 57701, USA}
\newcommand{\SouthDakotaState}{South Dakota State University, Brookings, SD 57007, USA}
\newcommand{\StonyBrook}{Stony Brook University, SUNY, Stony Brook, NY 11794, USA}
\newcommand{\SURF}{Sanford Underground Research Facility, Lead, SD, 57754, USA}
\newcommand{\Sussex}{University of Sussex, Brighton, BN1 9RH, United Kingdom}
\newcommand{\Syracuse}{Syracuse University, Syracuse, NY 13244, USA}
\newcommand{\Tecnologica }{Universidade Tecnol\'ogica Federal do Paran\'a, Curitiba, Brazil}
\newcommand{\TelAviv}{Tel Aviv University, Tel Aviv-Yafo, Israel}
\newcommand{\TexasAMcollege}{Texas A\&M University, College Station, Texas 77840}
\newcommand{\TexasAMcorpuscristi}{Texas A\&M University - Corpus Christi, Corpus Christi, TX 78412, USA}
\newcommand{\TexasArlington}{University of Texas at Arlington, Arlington, TX 76019, USA}
\newcommand{\Texasaustin}{University of Texas at Austin, Austin, TX 78712, USA}
\newcommand{\Toronto}{University of Toronto, Toronto, Ontario M5S 1A1, Canada}
\newcommand{\Tufts}{Tufts University, Medford, MA 02155, USA}
\newcommand{\Unifesp}{Universidade Federal de S\~ao Paulo, 09913-030, S\~ao Paulo, Brazil}
\newcommand{\UNIST}{Ulsan National Institute of Science and Technology, Ulsan 689-798, South Korea}
\newcommand{\UniversityCollegeLondon}{University College London, London, WC1E 6BT, United Kingdom}
\newcommand{\univkansas}{University of Kansas, Lawrence, KS 66045}
\newcommand{\UNMSM}{Universidad Nacional Mayor de San Marcos, Lima, Peru}
\newcommand{\ValleyCity}{Valley City State University, Valley City, ND 58072, USA}
\newcommand{\Vigo}{University of Vigo, E- 36310 Vigo Spain}
\newcommand{\VirginiaTech}{Virginia Tech, Blacksburg, VA 24060, USA}
\newcommand{\Warsaw}{University of Warsaw, 02-093 Warsaw, Poland}
\newcommand{\Warwick}{University of Warwick, Coventry CV4 7AL, United Kingdom}
\newcommand{\Wellesley}{Wellesley College, Wellesley, MA 02481, USA}
\newcommand{\Wichita}{Wichita State University, Wichita, KS 67260, USA}
\newcommand{\WilliamMary}{William and Mary, Williamsburg, VA 23187, USA}
\newcommand{\Wisconsin}{University of Wisconsin Madison, Madison, WI 53706, USA}
\newcommand{\Yale}{Yale University, New Haven, CT 06520, USA}
\newcommand{\Yerevan}{Yerevan Institute for Theoretical Physics and Modeling, Yerevan 0036, Armenia}
\newcommand{\York}{York University, Toronto M3J 1P3, Canada}
\affiliation{\Albanysuny}
\affiliation{\Almaty}
\affiliation{\Amsterdam}
\affiliation{\Antalya}
\affiliation{\Antananarivo}
\affiliation{\Antioquia}
\affiliation{\AntonioNarino}
\affiliation{\Argonne}
\affiliation{\Arizona}
\affiliation{\Asuncion}
\affiliation{\Athens}
\affiliation{\Atlantico}
\affiliation{\Augustana}
\affiliation{\Bern}
\affiliation{\Beykent}
\affiliation{\Birmingham}
\affiliation{\BolognaUniversity}
\affiliation{\Boston}
\affiliation{\Bristol}
\affiliation{\Brookhaven}
\affiliation{\Bucharest}
\affiliation{\CalBerkeley}
\affiliation{\CalDavis}
\affiliation{\CalIrvine}
\affiliation{\CalLosangeles}
\affiliation{\CalRiverside}
\affiliation{\CalSantabarbara}
\affiliation{\Caltech}
\affiliation{\Cambridge}
\affiliation{\Campinas}
\affiliation{\CataniaUniversitadi}
\affiliation{\Catolica}
\affiliation{\CBPF}
\affiliation{\CEASaclay}
\affiliation{\CERN}
\affiliation{\Charles}
\affiliation{\Chicago}
\affiliation{\ChungAng}
\affiliation{\CIEMAT}
\affiliation{\Cincinnati}
\affiliation{\Cinvestav}
\affiliation{\Colima}
\affiliation{\ColoradoBoulder}
\affiliation{\ColoradoState}
\affiliation{\Columbia}
\affiliation{\conida}
\affiliation{\Cti}
\affiliation{\CUSB}
\affiliation{\CzechAcademyofSciences}
\affiliation{\CzechTechnical}
\affiliation{\DannecyleVieux}
\affiliation{\Daresbury}
\affiliation{\Dordt}
\affiliation{\Drexel}
\affiliation{\Duke}
\affiliation{\Durham}
\affiliation{\Edinburgh}
\affiliation{\EIA}
\affiliation{\Eotvos}
\affiliation{\erciyes}
\affiliation{\FCULport}
\affiliation{\FederaldeAlfenas}
\affiliation{\FederaldeGoias}
\affiliation{\FederaldoABC}
\affiliation{\FederaldoRio}
\affiliation{\Fermi}
\affiliation{\Ferrarauniv}
\affiliation{\Florida}
\affiliation{\Floridastate}
\affiliation{\Fluminense}
\affiliation{\Genova}
\affiliation{\Georgian}
\affiliation{\Granada}
\affiliation{\GranSasso}
\affiliation{\GranSassoLab}
\affiliation{\Grenoble}
\affiliation{\Guanajuato}
\affiliation{\Harish}
\affiliation{\Hawaii}
\affiliation{\hkust}
\affiliation{\Houston}
\affiliation{\Hyderabad}
\affiliation{\Idaho}
\affiliation{\IFIC}
\affiliation{\IGFAE}
\affiliation{\ihep}
\affiliation{\Iitk}
\affiliation{\Illinoisinstitute}
\affiliation{\Imperial}
\affiliation{\IndGuwahati}
\affiliation{\IndHyderabad}
\affiliation{\Indiana}
\affiliation{\INFNBologna}
\affiliation{\INFNCatania}
\affiliation{\INFNFerrara}
\affiliation{\INFNFrascati}
\affiliation{\INFNGenova}
\affiliation{\INFNLecce}
\affiliation{\INFNMilanBicocca}
\affiliation{\INFNMilano}
\affiliation{\INFNNapoli}
\affiliation{\INFNPadova}
\affiliation{\INFNPavia}
\affiliation{\INFNPisa}
\affiliation{\INFNRoma}
\affiliation{\INFNRomavergata}
\affiliation{\INFNSud}
\affiliation{\Infntorino}
\affiliation{\Ingenieria}
\affiliation{\Insubria }
\affiliation{\Iowa}
\affiliation{\IowaState}
\affiliation{\IPLyon}
\affiliation{\IPM}
\affiliation{\IRLPPC}
\affiliation{\ISTlisboa}
\affiliation{\Ita}
\affiliation{\Iwate}
\affiliation{\Jacksonstate}
\affiliation{\Jawaharlal}
\affiliation{\Jeonbuk}
\affiliation{\Jyvaskyla}
\affiliation{\Kansasstate}
\affiliation{\Kavli}
\affiliation{\KEK}
\affiliation{\KISTI}
\affiliation{\Kyiv}
\affiliation{\Lancaster}
\affiliation{\LawrenceBerkeley}
\affiliation{\LIP}
\affiliation{\Liverpool}
\affiliation{\LosAlmos}
\affiliation{\Louisanastate}
\affiliation{\LpBordeaux}
\affiliation{\Lucknow}
\affiliation{\Mainz}
\affiliation{\Manchester}
\affiliation{\Massinsttech}
\affiliation{\Medellin}
\affiliation{\Michigan}
\affiliation{\Michiganstate}
\affiliation{\MilanoBicocca}
\affiliation{\MilanoUniv}
\affiliation{\Minnduluth}
\affiliation{\Minntwin}
\affiliation{\Mississippi}
\affiliation{\napoli}
\affiliation{\Nikhef}
\affiliation{\Niser}
\affiliation{\Northdakota}
\affiliation{\Northernillinois}
\affiliation{\Northwestern}
\affiliation{\NotreDame}
\affiliation{\NoviSad}
\affiliation{\Ohiostate}
\affiliation{\OregonState}
\affiliation{\Oxford}
\affiliation{\PacificNorthwest}
\affiliation{\Padova}
\affiliation{\Panjab}
\affiliation{\Parissaclay}
\affiliation{\Parisuniversite}
\affiliation{\Parma}
\affiliation{\Pavia}
\affiliation{\Penn}
\affiliation{\PennState}
\affiliation{\PhysicalResearchLaboratory}
\affiliation{\Pisa}
\affiliation{\Pitt}
\affiliation{\Pontificia}
\affiliation{\PuertoRico}
\affiliation{\Punjab}
\affiliation{\QMUL}
\affiliation{\Radboud}
\affiliation{\Rice}
\affiliation{\Rochester}
\affiliation{\Royalholloway}
\affiliation{\Rutgers}
\affiliation{\Rutherford}
\affiliation{\Salento}
\affiliation{\santamarta}
\affiliation{\Sapienza}
\affiliation{\SergioArboleda}
\affiliation{\Sheffield}
\affiliation{\SLAC}
\affiliation{\Southcarolina}
\affiliation{\SouthDakotaSchool}
\affiliation{\SouthDakotaState}
\affiliation{\StonyBrook}
\affiliation{\SURF}
\affiliation{\Sussex}
\affiliation{\Syracuse}
\affiliation{\Tecnologica }
\affiliation{\TelAviv}
\affiliation{\TexasAMcollege}
\affiliation{\TexasAMcorpuscristi}
\affiliation{\TexasArlington}
\affiliation{\Texasaustin}
\affiliation{\Toronto}
\affiliation{\Tufts}
\affiliation{\Unifesp}
\affiliation{\UNIST}
\affiliation{\UniversityCollegeLondon}
\affiliation{\univkansas}
\affiliation{\UNMSM}
\affiliation{\ValleyCity}
\affiliation{\Vigo}
\affiliation{\VirginiaTech}
\affiliation{\Warsaw}
\affiliation{\Warwick}
\affiliation{\Wellesley}
\affiliation{\Wichita}
\affiliation{\WilliamMary}
\affiliation{\Wisconsin}
\affiliation{\Yale}
\affiliation{\Yerevan}
\affiliation{\York}
\author{S.~Abbaslu} \affiliation{\IPM}
\author{F.~Abd Alrahman} \affiliation{\Houston}
\author{A.~Abed Abud} \affiliation{\CERN}
\author{R.~Acciarri} \affiliation{\CERN}
\author{L.~P.~Accorsi} \affiliation{\Tecnologica }
\author{M.~A.~Acero} \affiliation{\Atlantico}
\author{M.~R.~Adames} \affiliation{\Tecnologica }
\author{G.~Adamov} \affiliation{\Georgian}
\author{M.~Adamowski} \affiliation{\Fermi}
\author{C.~Adriano} \affiliation{\Campinas}
% ...
\collaboration{The DUNE Collaboration}
\noaffiliation

\begin{abstract}
\noindent
The ProtoDUNE-SP detector, a kiloton-scale prototype for the Deep Underground Neutrino Experiment (DUNE) far detector, is the largest liquid argon time projection chamber built to date. Operated at CERN from 2018 to 2020, it collected both cosmic-ray data and a beam consisting of positively-charged particles with discrete momentum settings across a range of 0.3 GeV/$c$ to 7 GeV/$c$. In this letter, we report the total inelastic cross section measurements for $\pi^+$--Ar and $p$--Ar interactions using selected $\pi^+$ and proton samples from the 1 GeV/$c$ beam data, spanning kinetic energies of 500--900~MeV and below 450~MeV, respectively. These energy ranges are directly relevant to hadrons produced in DUNE. The measured cross sections are consistent with predictions and provide a dataset that was previously unavailable for argon targets. These measurements are essential for constraining neutrino-argon interaction models and achieving the precision physics goals of the upcoming DUNE experiment.

\end{abstract}

\maketitle

\setlength{\parskip}{0.5pt}
\modulolinenumbers[1]
The Deep Underground Neutrino Experiment (DUNE)~\cite{DUNEtdr1,DUNE} is a next-generation neutrino oscillation experiment designed to determine the neutrino mass ordering and measure charge-parity (CP) violation in the lepton sector. DUNE consists of a near detector complex, located close to the source of a high-intensity neutrino beam produced by the Long-Baseline Neutrino Facility~\cite{DUNEtdr1}, and a far detector located 1300 kilometers from the source. The detectors' design centers on the liquid argon time projection chamber (LArTPC) technology, which offers detailed 3D tracking of charged particles produced in neutrino interactions with argon. The far detector will contain up to 70 kilotons of liquid argon in total. To validate the scalability of the technology to this unprecedented size, prototype detectors have been developed and constructed. As the first of these prototypes, ProtoDUNE Single-Phase (ProtoDUNE-SP)~\cite{ProtoDUNE_construction} successfully operated at the CERN Neutrino Platform\mbox{~\cite{Pietropaolo:2017jlh}} from 2018 to 2020, demonstrating high performance with a 770-ton LArTPC~\cite{ProtoDUNE-SP}. 
In addition to recording cosmic-ray data, ProtoDUNE-SP was exposed in autumn 2018 to a positively-charged particle beam, including pions, protons, and kaons, with momentum settings of 0.3, 0.5, 1, 2, 3, 6, and 7 GeV/$c$~\cite{Charitonidis:2017omo,Booth:2019brj}. This beam exposure provides a dedicated dataset to characterize hadron-argon interactions, which are essential for reducing interaction-related uncertainties in neutrino energy reconstruction for DUNE and other LArTPC-based neutrino experiments\mbox{~\cite{MicroBooNE,SBND,ICARUS}}.

DUNE’s beam will deliver neutrinos with energies up to several GeV, where the majority of hadrons produced in initial neutrino interactions are nucleons and charged pions, whose kinetic energies typically peak around a few hundred MeV and extend beyond 1~GeV~\cite{DUNE}. In this regime, final-state interactions (FSI), where the produced hadrons further scatter within the nucleus before exiting, significantly distort final-state kinematics and even change the types of detectable particles. Mismodeling these effects introduces substantial uncertainties in event reconstruction, which can bias the neutrino energy measurement and potentially obscure sensitivity to oscillation parameters, most notably the CP-violating phase~\cite{Dytman:2021ohr,Liu:2025hpl}. 
Additionally, secondary interactions of the emitted hadrons as they propagate through the liquid argon contribute to the complexity of event reconstruction~\cite{Friedland:2018vry}. Accurate modeling of both FSI and secondary interactions requires hadron-argon cross section data, yet such data remain scarce, forcing predictions to rely on interpolation from data on solid targets such as carbon and lead~\cite{PinzonGuerra:2018rju,Carlson:1996ofz}. In 1999, the LADS experiment~\cite{Backenstoss:1991ej} reported the first measured $\pi^+$ absorption cross section on argon at three discrete pion kinetic energies: 118, 162, and 239 MeV~\cite{LADS}. Additional experimental data did not appear until 2021, when the LArIAT collaboration~\cite{LArIAT:2019kzd} measured the $\pi^-$--argon total hadronic cross section for $\pi^-$ kinetic energies below 700 MeV~\cite{LArIAT}. In 2024, ProtoDUNE-SP presented the first measurement of kaon-argon interactions using 6 and 7 GeV/$c$ beam data\mbox{~\cite{DUNE:2024qgl}}. This work, contributing to the same experimental program, presents the first measurement of total inelastic $\pi^+$--Ar and $p$--Ar cross sections in the kinetic energy ranges of 500-900 MeV and below 450 MeV, respectively, filling a longstanding gap in argon data and benchmarking hadronic interaction models for the global LArTPC neutrino program. 

The ProtoDUNE-SP detector~\cite{DUNE:2017pqt} consists of two TPCs, separated by a shared cathode plane located at the center, defined at $x=0$. The $y$ axis is vertical, and the beam is roughly aligned with the positive $z$ direction, deviating by approximately 15$^\circ$ towards the negative $x$ and negative $y$ directions. The beam enters the LArTPC on the negative $x$ side through its front face, where $z=0$ is defined. At the entry point, a nitrogen-filled beam plug~\cite{ProtoDUNE_construction} is installed to minimize the energy loss of beam particles compared to direct entry through the cryostat materials. On both the beam side ($x<0$) and the non-beam side ($x>0$), three anode plane assemblies (APAs) are installed in parallel. Each APA contains three layers of sensing wires oriented at $35.7^\circ$, $-35.7^\circ$, and $0^\circ$ from vertical, corresponding to two induction planes (U, V) and one collection plane (X). In a LArTPC detector, when charged particles travel through the liquid argon, they ionize the surrounding argon atoms. The resulting ionization electrons drift toward the anode plane under a nominal electric field of 500 V/cm. These drifting electrons induce currents on the sensing wires as they pass through the U and V planes and are collected on the X plane. The different orientations of the wire layers allow for the reconstruction of a 2D projection of the charged particle tracks. The time difference between the charge signal on each wire and the event's start time, $t_0$, is used to determine the distance from the anode plane, enabling 3D event reconstruction. For beam particles, $t_0$ is provided by a trigger signal from the beam line instrumentation~\cite{ProtoDUNE-SP}.

The beam line instrumentation provides time-of-flight (TOF) information and Cherenkov detectors 
for beam particle identification (PID)~\cite{ProtoDUNE-SP}. Using its 1 GeV/$c$ beam trigger, ProtoDUNE-SP collected a data sample with over 400,000 events. The 1 GeV/$c$ beam has a momentum spread of over 60 MeV/$c$ with a resolution of approximately 40 MeV/$c$ as measured by the beam line instrumentation. The beam is simulated using \texttt{G4beamline}~\cite{Roberts:2007nte}. 
Particle transportation and interactions inside the detector are handled by \textsc{Geant4}~\cite{GEANT4:2002zbu,Allison:2006ve,Allison:2016lfl} with the LArSoft~\cite{Church:2013hea} toolkit. Event reconstruction is performed using the Pandora software package~\cite{Marshall:2015rfa} customized for ProtoDUNE-SP~\cite{DUNE:2022wlc}. Pandora utilizes pattern recognition to cluster energy depositions on each wire plane, and reconstructs these clusters across all three planes into tracks and showers. Subsequently, Pandora attempts to identify the beam cluster in each beam-triggered event. Located on the surface, ProtoDUNE-SP is exposed to a considerable flux of cosmic rays, which results in the accumulation of slowly drifting positive ions. This is the so-called ``space-charge effect", which distorts the electric field and affects reconstruction. A data-driven simulation of the effect was produced~\cite{SCE}, and the derived space-charge map is applied as a correction to all reconstructed variables measured by Pandora and downstream analyses.

\begin{figure}[htbp]
    \centering
    \includegraphics[width=0.45\textwidth]{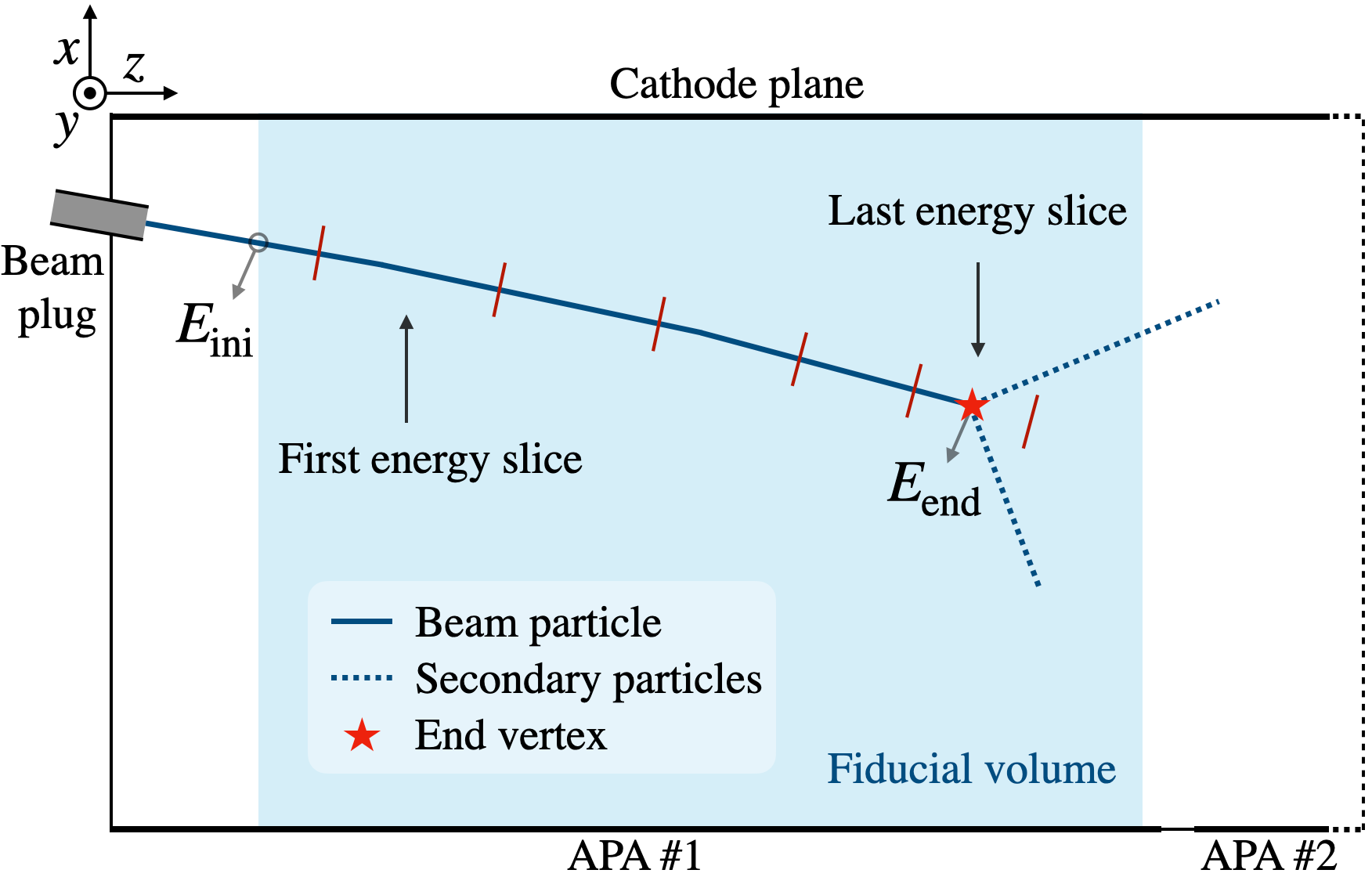}
    \caption{A top-down view illustrating the slicing method, which shows a portion of the ProtoDUNE-SP detector on the beam side within the range of the first APA.
    } 
    \label{fig:slicing_method}
\end{figure}

In traditional hadron-nucleus cross section measurements, a thin target of the material is typically used. However, a LArTPC is not a thin target. Its size far exceeds the mean free path of a hadron in liquid argon in this energy regime, which is on the order of 10 cm. To address this, the LArIAT collaboration proposed the thin-slice method~\cite{LArIAT}, where the detector is hypothetically divided into several thin slices, with each slice functioning as an individual thin-target experiment. The interaction location and energy of the beam particle in each thin slice are determinable thanks to the granularity and accurate event reconstruction of LArTPCs. In this work, a modified version of the slicing method~\cite{Liu:2023psg} is used, where beam tracks are divided into energy slices rather than spatial slices. 
As illustrated in Fig.~\ref{fig:slicing_method}, the beam particle enters the detector fiducial volume with an initial kinetic energy ($E_{\rm ini}$) and continuously loses energy while traversing the liquid argon, a process described by the Bethe-Bloch formula~\cite{PDG}. Its kinetic energy at the end vertex is denoted $E_{\rm end}$. With this method, the total inelastic cross section, $\sigma(E)$, is calculated using the following formula: 
\begin{equation}
    \sigma(E)=\frac{N_{\rm int}(E)}{nN_{\rm end}(E)\delta E}\frac{dE}{dx}(E)\ln\left(\frac{N_{\rm inc}(E)}{N_{\rm inc}(E)-N_{\rm end}(E)}\right).
    \label{eqn:cross-section}
\end{equation}
Here, $N_{\rm end}$ is the distribution of $E_{\rm end}$, while $N_{\rm inc}$ and $N_{\rm int}$ indicate the number of incident and interacting particles in a given energy slice of width $\delta E$, and $n$ is the argon number density. The detailed definitions of these energy histograms ($N_{\rm ini}$, $N_{\rm end}$, $N_{\rm int}$, $N_{\rm inc}$), the reconstruction of the energies ($E_{\rm ini}$ and $E_{\rm end}$), and the full derivation of the formula are provided as the Supplemental Material.


The analysis is performed separately for two beam samples, identified by the beam line PID as $\pi^+$/$\mu^+$ and protons~\cite{ProtoDUNE-SP}. The signal events are defined as the $\pi^+$ or proton inelastic interaction events, respectively, with no constraints on scattering angles and final-state topology. Further selections are applied to the Pandora-identified beam track based on the reconstructed information in the LArTPC. 
A pre-selection criterion ensures sufficient charge deposition on the wire planes to indicate beam activity within the detector. Additional constraints are then imposed on the beam entrance position and the beam angle. The fiducial volume for ProtoDUNE-SP is defined as $z \in [30, 220]$ cm, and only track segments within this range are considered. The lower boundary removes beam tracks shorter than 30 cm, whose identification efficiency is non-uniform and difficult to model in simulation. The upper boundary addresses distortions in the electric field caused by an unintentionally grounded electron diverter, a component designed to modify the local drift field. This issue led to reconstructed tracks being broken near the gap between the first two APAs on the beam side~\cite{ProtoDUNE-SP}. As simulation suggests fewer than 10\% of 1 GeV/$c$ pion tracks and no 1 GeV/$c$ proton tracks extend into the region of the second APA, the upper boundary selection, $z < 220$ cm, is applied right before the gap. 

Following the selection criteria shared by both the beam $\pi^+$/$\mu^+$ and proton samples, specific vetoes against major background components are applied. While the beam muons cannot be discriminated from the pions using the beam line PID given their similar mass, the fiducial volume criterion helps mitigate the muon background, as most 1 GeV/$c$ muon tracks reach the upper boundary. An additional criterion suppresses muon backgrounds by determining whether the identified beam track ends in a Michel electron, a feature for stopping muons. A Michel score based on a convolutional neural network~\cite{DUNE:2022fiy} trained on simulations is evaluated for each identified beam track to facilitate this selection. Secondary protons, which are products of beam interactions but misidentified as the beam particles, are another major background for beam pions. A $\chi^2$ value calculated against the proton stopping power profile~\cite{ProtoDUNE-SP} is employed to reduce this background. In the proton analysis, stopping protons, which come to rest without interacting inelastically with an argon nucleus, are distinguished from proton inelastic events using the $\chi^2$ on proton stopping power, as well as the continuous-slowing-down-approximation (CSDA)~\cite{groom2012muon} given the proton energy. According to simulations, over 99\% of the elastic scatters in this energy regime occurs at angles smaller than 3$^\circ$, with no vertex reconstructed by Pandora. In contrast, inelastic scattering vertices with larger scattering angles and more final-state particles are often detected. Consequently, elastic backgrounds are fully estimated using Monte Carlo (MC) samples, and no angle constraint is applied to define the inelastic signal.

During the selection process, discrepancies between data and MC are corrected by reweighting the MC sample for variables not directly entering the cross section extraction. In the pion analysis, the fraction of beam muons is found to be underestimated in the simulation, leading to fewer long tracks in MC compared to data. A $\chi^2$ fit to the reconstructed track length distribution for tracks longer than 150 cm---which is dominated by tracks extending beyond the fiducial volume---yields a scale factor of $1.6 \pm 0.2$, which is applied to MC beam muons. 
Additionally, beam momentum spectra in MC are found to be narrower than those in data, by 17\% for beam $\pi^+$/$\mu^+$ and 11\% for beam protons. These are corrected using stopping-particle samples: a Michel-tagged muon sample in the pion analysis and a sideband-selected stopping proton sample in the proton analysis. In both cases, the beam momentum is estimated from the reconstructed track length, and reweighting parameters are obtained by fitting the length-derived momentum distributions.

\begin{figure*}[htbp]
    \centering
    \includegraphics[width=0.45\textwidth]{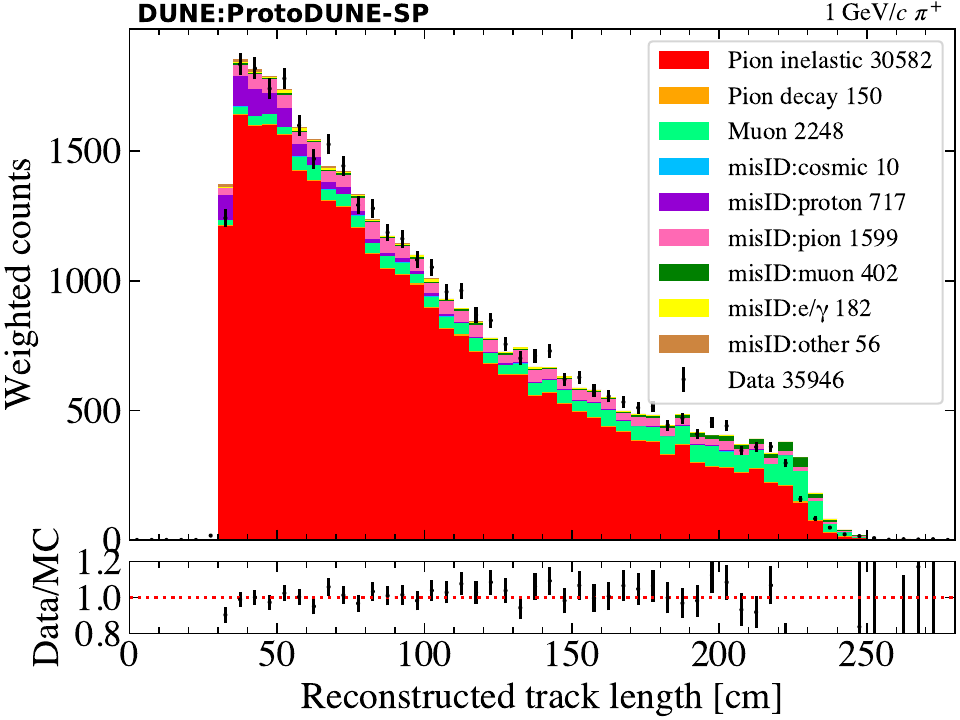}\put(-215,165){(a)}
    \includegraphics[width=0.45\textwidth]{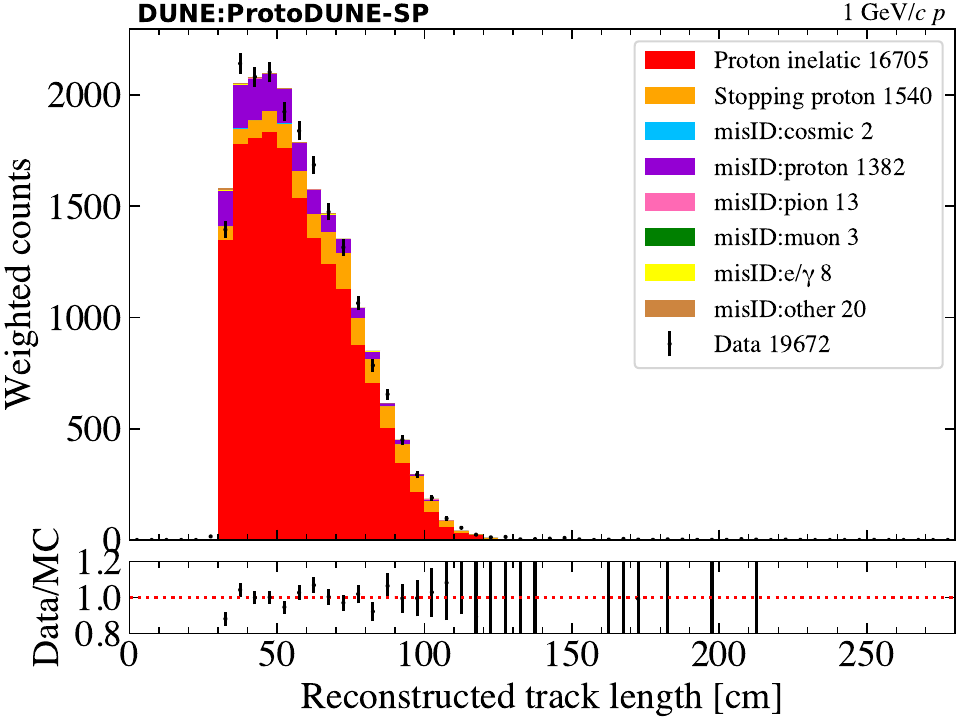}\put(-215,165){(b)}
    \caption{Reconstructed track length distributions after full selection for the (a) pion and (b) proton analyses. Error bars on data points are statistical-only. The stacked histograms show different event categories from the MC, with the total weighted counts normalized to match the data. The numbers in the legend indicate the weighted counts for each component. The bottom panel in each subplot displays the ratio of data to MC in each bin. MC reweighting, as described in the main text, is applied.}
    \label{fig:after_full_selection}
\end{figure*}

After full selection, the efficiency and purity for pion inelastic events are estimated by the reweighted MC to be 64\% and 85\%, respectively; for proton inelastic events, they are 53\% and 85\%.
Figure~\ref{fig:after_full_selection} shows the reconstructed track length distributions after full selection for both the pion and proton analyses, where the MC sample is displayed as stacked histograms, with signal (in red) and background components shown separately.
Remaining backgrounds are subtracted from data histograms using MC estimations. For major background components with fractions exceeding 1\%, data--MC differences are evaluated in sideband regions where the respective background dominates and the distribution is relatively flat. According to MC estimates, the selected pion sample contains approximately 7.3\% muons (including 6.3\% reweighted beam muons and 1.0\% secondary muons), 4.4\% secondary pions, and 2.0\% secondary protons. A fit to the muon-enriched sideband of the Michel score distribution yields a scale factor of $0.93 \pm 0.12$ for muons. For secondary protons, 
a scale factor of $1.7 \pm 0.2$ is obtained by fitting the proton $\chi^2$ distribution in the sideband region. The secondary pion background is similarly constrained using the beam angle distribution, yielding a scale factor of $1.3 \pm 0.2$. In the proton analysis, the main backgrounds include 7.8\% stopping protons and 7.0\% secondary protons. Sideband fits to the proton $\chi^2$ distribution and the beam angle distribution indicate that no additional scaling is required for either component.

Detector efficiency and resolution effects need to be corrected for following selection and background subtraction for the energy histograms to be used for cross section calculation. While the original thin-slice method applies bin-by-bin corrections~\cite{LArIAT}, this analysis models the detector response between true and reconstructed values using the MC sample and applies an unfolding procedure~\cite{unfolding}. Given the fact that the energy histograms are not independent, a multi-dimensional unfolding is performed to account for full correlations, where the indices of $N_{\rm ini}$, $N_{\rm end}$, and $N_{\rm int}$ are flattened into a single variable, denoted $x$, for unfolding. The D'Agostini method~\cite{DAgostini} based on the iterative Bayesian unfolding~\cite{Richardson:1972hli,Lucy:1974yx}, implemented in the RooUnfold package~\cite{RooUnfold}, is used. It contains one regularization parameter: the number of iterations, $n$. In general, unfolded results with smaller $n$ are more biased toward the input MC, while larger $n$ bring larger variations. The optimization of the regularization parameter is guided by a data-driven test, where a statistic,
$t=(x_n-x_{\rm unreg})\cdot (V_n+V_{\rm unreg})^{-1}\cdot(x_n-x_{\rm unreg})$, is constructed to evaluate the consistency between the unfolded histogram $x_n$ at a given $n$ and the unregularized result $x_{\rm unreg}$, defined as the limit obtained with an infinite number of iterations in the D'Agostini unfolding. $V$ denotes the corresponding covariance matrices. The test statistic $t$ decreases with $n$, reflecting reduced bias, and the smallest $n$ yielding a threshold value is chosen for the nominal result. 
The optimal number of iterations is determined to be $n = 47$ for the pion analysis and $n = 16$ for the proton analysis. Nonetheless, the results are not sensitive to the precise value of $n$.


As data statistical uncertainty is propagated analytically from the output of unfolding, the systematic uncertainties associated with the MC are treated in the same manner in both the pion and proton analyses. These uncertainties impact the analysis by providing models for background shapes, the response matrix, and the efficiency across all bins. The systematic uncertainty associated with background estimations is treated by including the statistical fluctuations of background histograms and the fit errors of the scale factors. Systematic uncertainties associated with finite MC sample size, cross section models, energy reconstruction, and momentum reweighting are treated by generating pseudo-experiments with fluctuated parameters related to each source according to their uncertainties. For the MC sample size, the response matrix and the efficiency plot are fluctuated according to the statistical error in each bin. The uncertainty for the total inelastic cross section model is estimated as an overall scale of $\pm 5\%$ given the general agreement between models and measurements on other targets~\cite{PinzonGuerra:2018rju}. The uncertainty for the reconstructed energies is estimated as a constant 4 MeV for pion and 2 MeV for proton, conservatively derived from the Gaussian fit uncertainties to the energy distributions. For the momentum reweighting, the parameters are fluctuated according to their fit errors. To estimate the uncertainty on space-charge corrections, an alternative space-charge map based on simulation is produced, and the difference between results derived from both maps are estimated as its systematic uncertainty. Uncertainties on factors appearing in the cross section formula, such as argon density and the $dE/dx$ values in each energy bin, are also accounted for by varying their nominal values. Among all sources, limited MC sample size is found to dominate the total systematic uncertainty, while background modeling and energy reconstruction contribute more substantially in the higher energy region. The total covariance matrix is obtained by summing those corresponding to individual uncertainties.

\begin{figure}[htbp]
    \centering
    \includegraphics[width=0.45\textwidth]{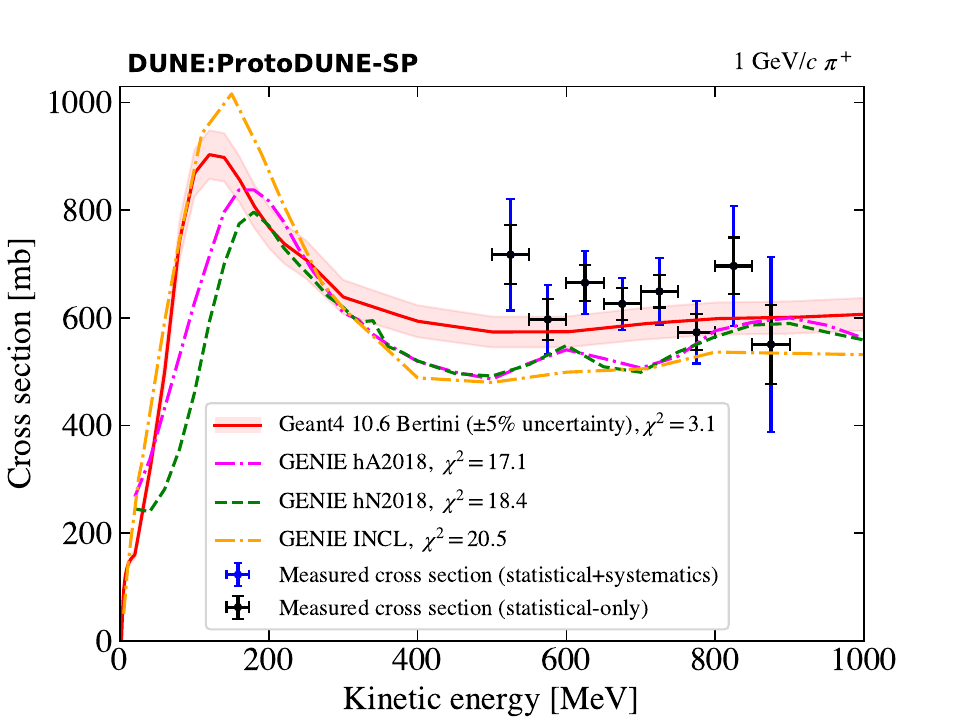}\put(-230,138){(a)}
    
    \includegraphics[width=0.45\textwidth]{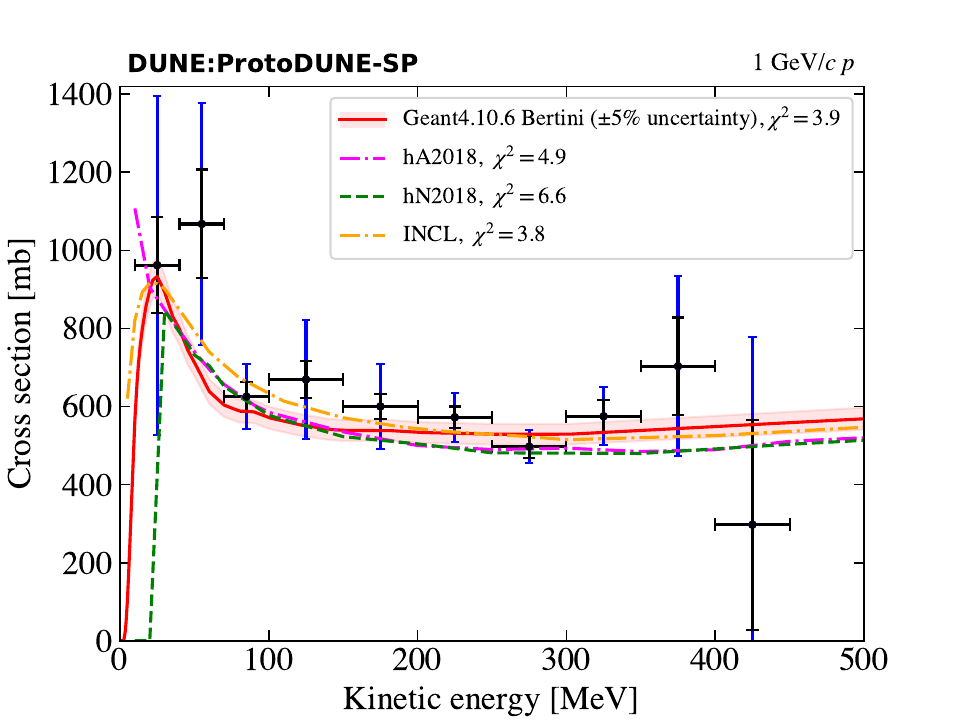}\put(-230,138){(b)}
    \caption{The measured (a) $\pi^+$--Ar and (b) $p$--Ar cross sections. In each sub-figure, the red solid curve indicates the cross section model used in simulation, with the shaded band indicating the fluctuation considered in the systematic uncertainty for the cross section model. Other models are also shown. The correlation matrices with all uncertainties applied are provided as part of the Supplemental Material.}
    \label{fig:XS_model}
\end{figure}

\begin{figure}[htbp]
    \centering
    \includegraphics[width=0.45\textwidth]{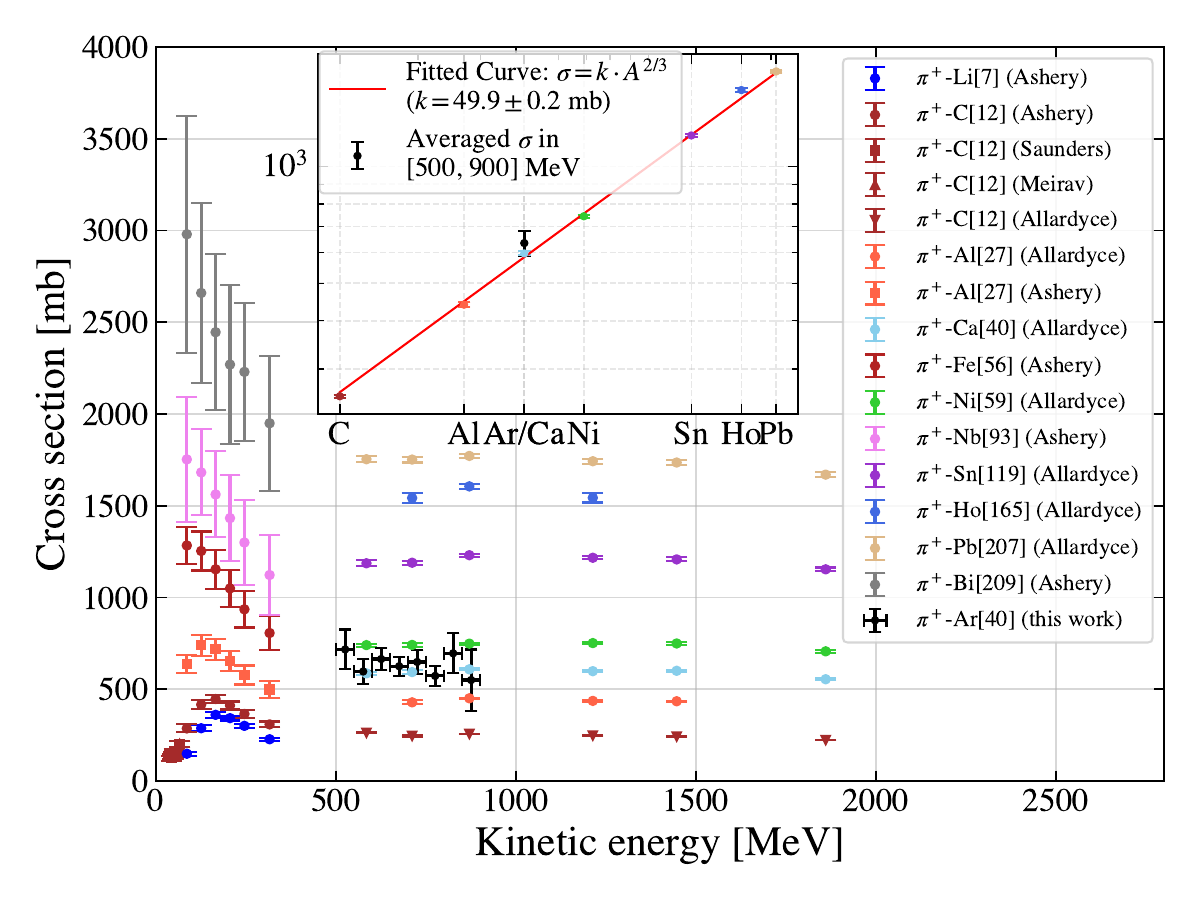}\put(-225,150){(a)}
    
    \includegraphics[width=0.45\textwidth]{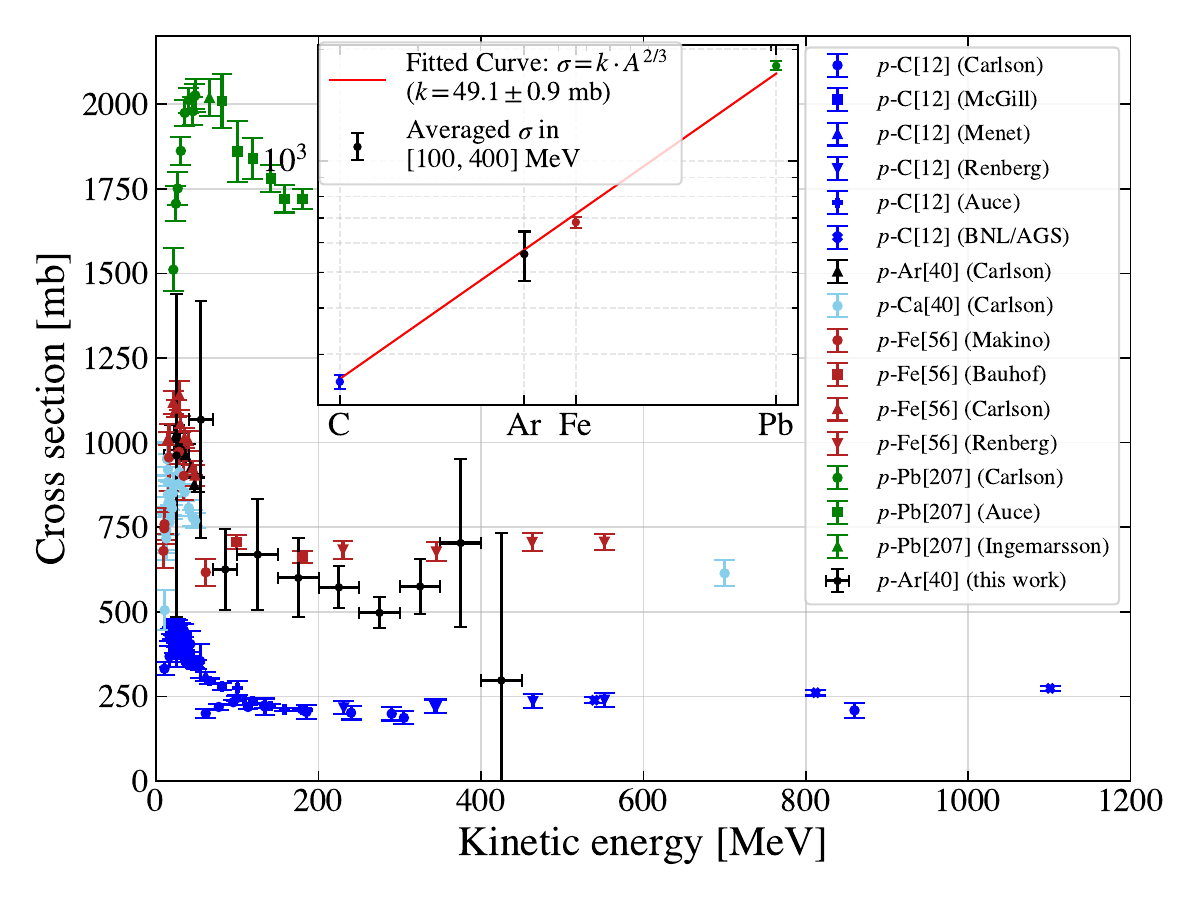}\put(-226,150){(b)}
    \caption{The cross section measurements on various targets with (a) $\pi^+$ and (b) proton. Our argon results are shown as black points. In the inner panel in each sub-figure, an empirical curve $\sigma=k\cdot A^{2/3}$ is fitted to the datasets averaged in the kinetic energy ranges [500, 900] MeV for $\pi^+$ and [100, 400] MeV for protons.}
    \label{fig:XS-A}
\end{figure}

Figure~\ref{fig:XS_model} presents the measured cross sections for the pion and proton data samples, respectively. 
The cross section model used in the simulation (\textsc{Geant4} 10.6 Bertini~\cite{Guthrie:1968ue,Bertini:1971xb,Wright:2015xia}, implemented as the QGSP\_BERT physics list in LArSoft~\cite{Church:2013hea}) along with several other models\mbox{~\cite{Andreopoulos:2015wxa,Dytman:2021ohr}} are overlaid for visual comparison with the data results. The peak around 165 MeV in the pion curve corresponds to the $\Delta(1232)$ resonance, whereas the proton curve peaks around 30 MeV and then decreases due to compound nuclear processes~\cite{Dytman:2021ohr}. 
The $\chi^2/N_{\rm dof}$ values calculated for the data compared to the \textsc{Geant4} 10.6 total inelastic cross section model are $3.1/8$ and $3.9/10$, respectively. While other models yield larger $\chi^2$ for the pion analysis, they cannot be ruled out given the current uncertainties.
Figure~\ref{fig:XS-A} shows the measured cross sections overlaid with previous measurements on various targets~\cite{PinzonGuerra:2018rju,Carlson:1996ofz}. An empirical relation between the cross section $\sigma$ and the atomic mass number $A$ exists: $\sigma\propto A^{2/3}$~\cite{Carlson:1996ofz}. In the inner panels, our results on argon nuclei are shown, which also align with the relation.

In conclusion, we report the $\pi^+$--Ar and $p$--Ar total inelastic cross sections using 1~GeV/$c$ beam data collected by the ProtoDUNE-SP detector. The kinetic energy ranges of [500, 900]~MeV for $\pi^+$ and [10, 450]~MeV for protons span a range sensitive to FSI modeling, covering roughly 30\% and 60\% of the respective final-state hadron populations at DUNE. 
The results are consistent with the \textsc{Geant4} 10.6 total inelastic cross section model used in the simulation, providing the first dedicated argon measurements of $\pi^+$ and proton scattering for the LArTPC neutrino community. While uncertainties remain larger than those in measurements on other nuclear targets, these results provide a vital experimental benchmark beyond the reliance on interpolations from solid targets, and represent a key step toward achieving the precision required for oscillation measurements at DUNE, including the determination of the CP-violating phase. Additional work using the ProtoDUNE-SP dataset includes a separate analysis focused on exclusive cross sections~\cite{DUNE:2025zhx}, with further work underway on comprehensive differential cross sections. Furthermore, the methodology established in this analysis facilitates measurements with new datasets, particularly from the second phase of ProtoDUNE-SP, known as the ProtoDUNE Horizontal Drift (ProtoDUNE-HD)
. Featuring a similar configuration with both beam polarities and the ability to probe lower pion energies, ProtoDUNE-HD collected data in 2024 with approximately twice the statistics and strategically expanded simulation samples to mitigate leading systematics, aiming to extend these measurements to a broader phase space. Collectively, these measurements offer unique inputs for testing and tuning models of both hadronic FSI and secondary interactions in argon, thereby constraining systematic uncertainties that limit neutrino oscillation analyses. This expanding suite of hadron-argon scattering data establishes the foundational constraints necessary to support the physics goals of DUNE and the broader neutrino community.

\setlength{\parskip}{5pt}
The ProtoDUNE-SP detector was constructed and operated on the CERN Neutrino Platform.
We gratefully acknowledge the support of the CERN management, and the
CERN EP, BE, TE, EN and IT Departments for NP04/Proto\-DUNE-SP. This document was prepared by DUNE collaboration using the resources of the Fermi National Accelerator Laboratory (Fermilab), a U.S. Department of Energy, Office of Science, Office of High Energy Physics HEP User Facility. Fermilab is managed by Fermi Forward Discovery Group, LLC, acting under Contract No. 89243024CSC000002. This work was supported by
CNPq,
FAPERJ,
FAPEG, 
FAPESP and,
Fundação Araucária,             Brazil;
CFI, 
IPP and 
NSERC,                          Canada;
CERN;
ANID-FONDECYT,                  Chile;
M\v{S}MT,                       Czech Republic;
ERDF, FSE+,
Horizon Europe, 
MSCA and NextGenerationEU,      European Union;
CNRS/IN2P3 and
CEA,                            France;
PRISMA+,                        Germany;
INFN,                           Italy;
FCT,                            Portugal;
CERN-RO/CDI,                        Romania;
NRF,                            South Korea;
Generalitat Valenciana, 
Junta de Andalucía, 
MICINN, and 
Xunta de Galicia,               Spain;
SERI and 
SNSF,                           Switzerland;
T\"UB\.ITAK,                    Turkey;
The Royal Society and 
UKRI/STFC,                      United Kingdom;
DOE and 
NSF,                            United States of America.


\clearpage
\section{Supplemental Material}
The analysis employs a modified energy slicing method (illustrated in Fig.~\ref{fig:slicing_method} of the main text) that divides the beam particle track into energy slices based on its kinetic energies at the entry ($E_{\rm ini}$) and the end ($E_{\rm end}$). The calculation of the cross section relies on the following energy histograms:
\begin{itemize}
    \item The initial histogram ($N_{\rm ini}$) contains the first energy slice, corresponding to the near boundary of the fiducial volume.
    \item The end histogram ($N_{\rm end}$) contains the last energy slice, corresponding to the end vertex of the beam particle before it exits the far boundary of the fiducial volume.
    \item The interaction histogram ($N_{\rm int}$) contains the last energy slice of the beam particle which scatters inelastically.
    \item The incident histogram ($N_{\rm inc}$) represents the number of particles incident upon the energy slice, and is calculated from $N_{\rm ini}$ and $N_{\rm end}$.
\end{itemize}
The cross section formula Eq.~\ref{eqn:cross-section} depends on the number density of liquid argon ($n$), the energy bin width ($\delta E$), and the stopping power $dE/dx$ described by the Bethe-Bloch formula~\cite{PDG}. A full derivation of the cross section formula can be found in Ref.~\cite{Liu:2023psg}.

To measure $E_{\rm ini}$ and $E_{\rm end}$, the energy reconstruction is studied for three components separately: the initial kinetic energy of the beam particle $E_{\rm beam}$, the upstream energy loss $E_{\rm loss}$, and the energy deposition in the LArTPC $E_{\rm depo}$. The $E_{\rm beam}$ is derived from the momentum of the beam particle, which is measured by the beam line instrumentation. An analysis of the materials upstream of the detector using the beam line simulation determines the $E_{\rm loss}$, which is modeled by a second-order polynomial function of $E_{\rm beam}$. The beam energy at the front face of the LArTPC, denoted as $E_{\rm ff}$, is then given by $E_{\rm beam} - E_{\rm loss}$. Lastly, the energy deposition $E_{\rm depo}$ within the detector is estimated as the integral of $dE/dx$ along the reconstructed track length $L$. With these notations, $E_{\rm ini}$ is calculated as $E_{\rm ff} - E_{{\rm depo}}|_{L=L_{\rm ini}}$, where $L_{\rm ini}$ represents the track segment from the detector front face to the near boundary of the fiducial volume. The same expression applies to $E_{\rm end}$, with $L=L_{\rm end}$ defined to represent the full track length or its truncation at the far boundary of the fiducial volume.

\begin{figure*}[htbp]
    \centering
    \includegraphics[width=0.45\textwidth]{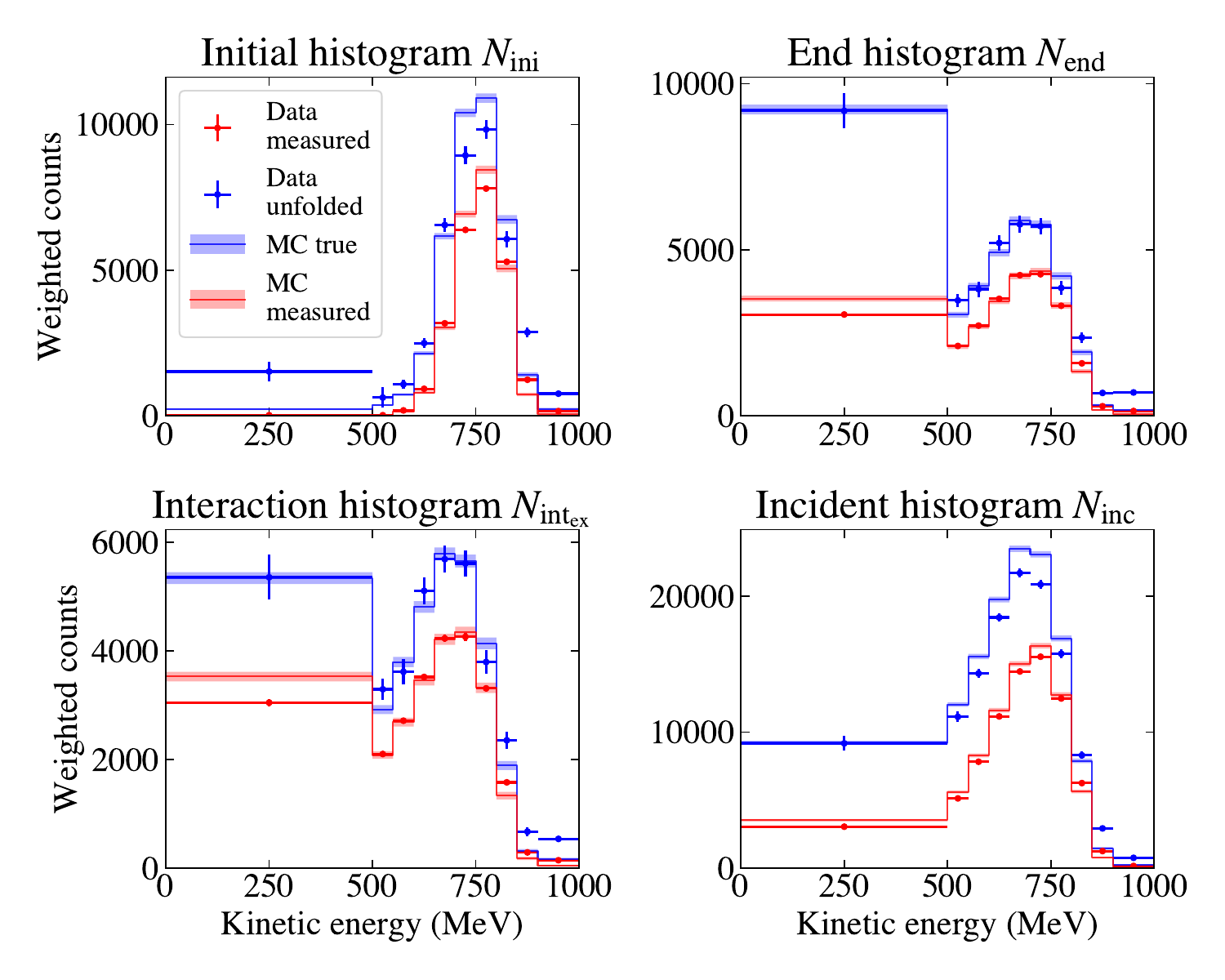}\put(-220,170){(a)}
    \includegraphics[width=0.45\textwidth]{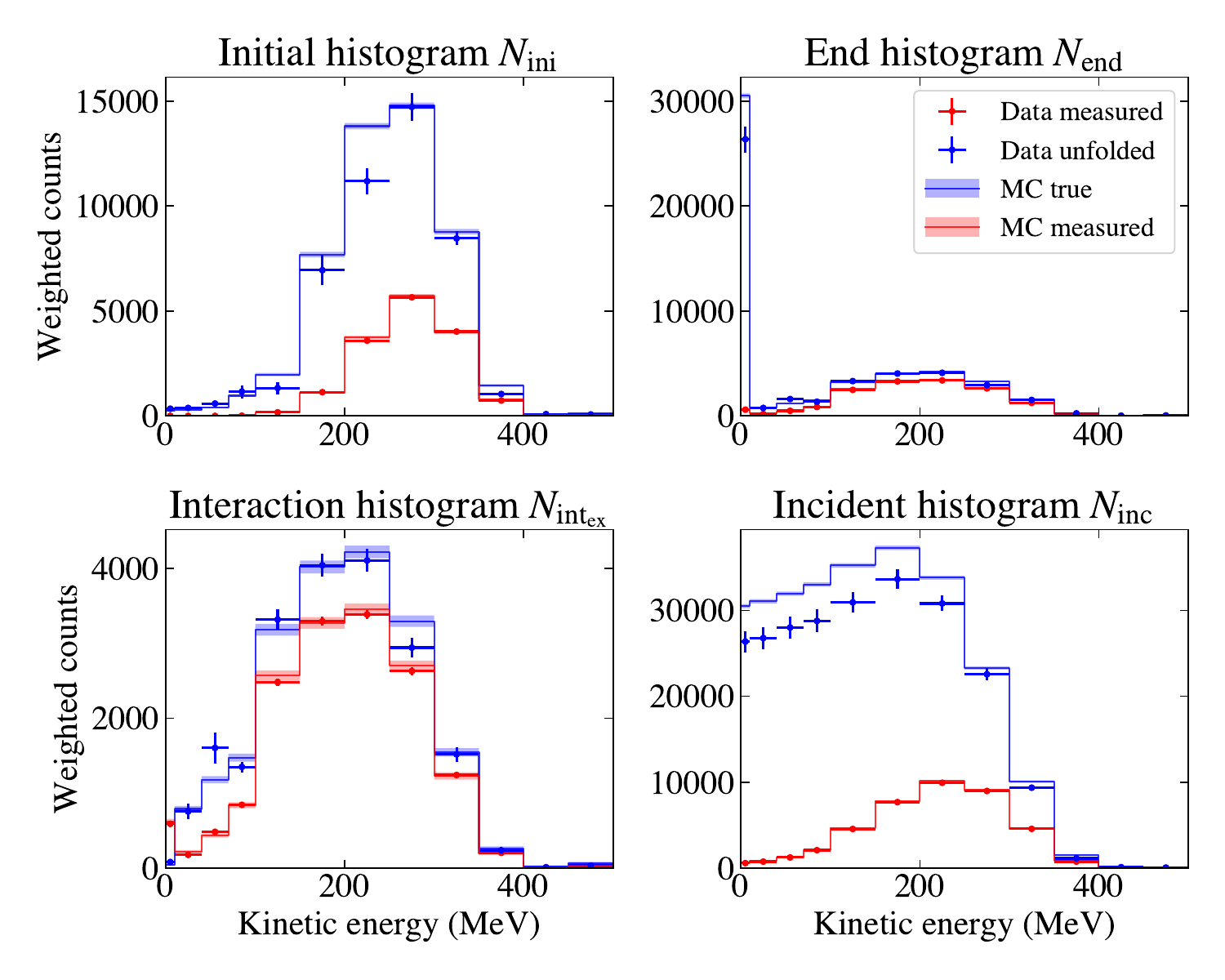}\put(-220,170){(b)}
    \caption{The energy histograms related to the cross section calculation ($N_{\rm ini}$, $N_{\rm end}$, $N_{\rm int}$, and $N_{\rm inc}$) before (red) and after (blue) unfolding for the (a) pion and (b) proton analyses. The error bars are statistical-only. MC is normalized to the data sample size before unfolding.}
    \label{fig:energy_histograms}
\end{figure*}

Following event selection and background subtraction, the energy histograms are corrected for detector effects via a multi-dimensional unfolding procedure as described in the main text. The data inputs for unfolding include the histogram of the flattened variable $x$ and the covariance matrix describing its statistical uncertainties. With the detector response estimated from the MC, the unfolding software (RooUnfold~\cite{RooUnfold}) outputs the unfolded histogram of $x$ and its corresponding covariance matrix, which is propagated to the covariance matrix for the final cross section results analytically\mbox{~\cite{Liu:2023psg}}. Figure~\ref{fig:energy_histograms} shows the energy histograms relevant to Eq.~\ref{eqn:cross-section} for both data and MC, illustrating the effects of the multi-dimensional unfolding projected onto the three energy histograms: $N_{\rm ini}$, $N_{\rm end}$, and $N_{\rm int}$, as well as the derived $N_{\rm inc}$.

\begin{figure*}[htbp]
    \centering
    \includegraphics[width=0.39\textwidth]{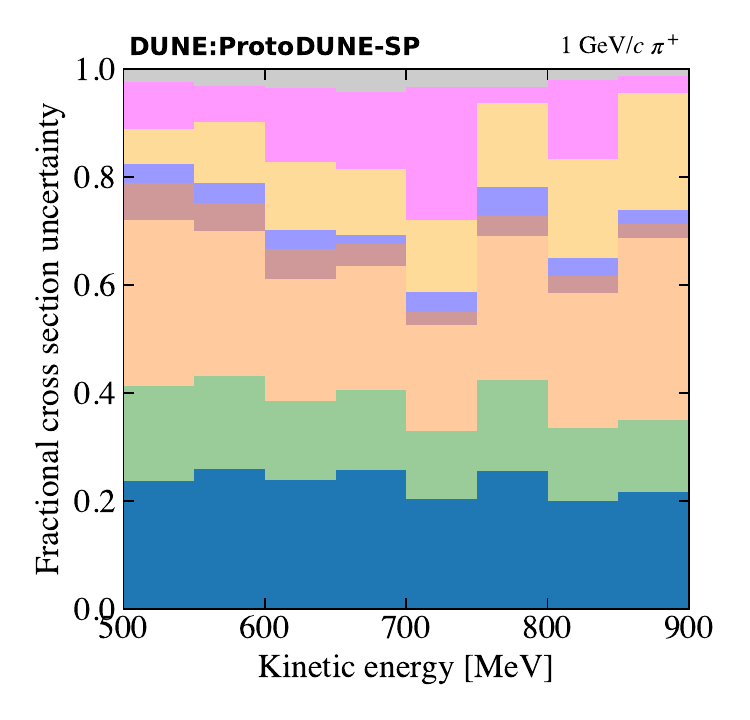}\put(-193,170){(a)}
    \includegraphics[width=0.21\textwidth]{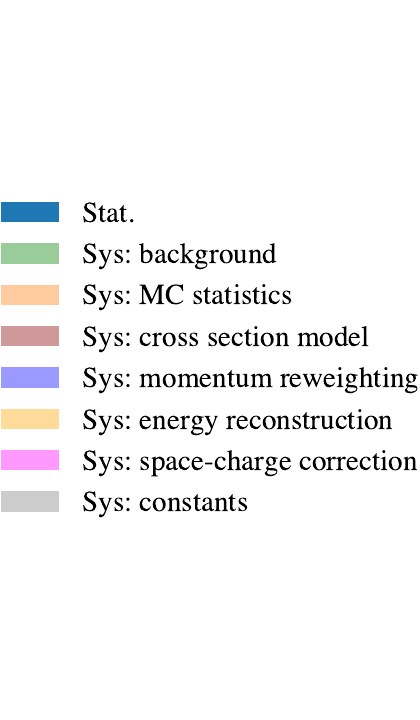}
    \includegraphics[width=0.39\textwidth]{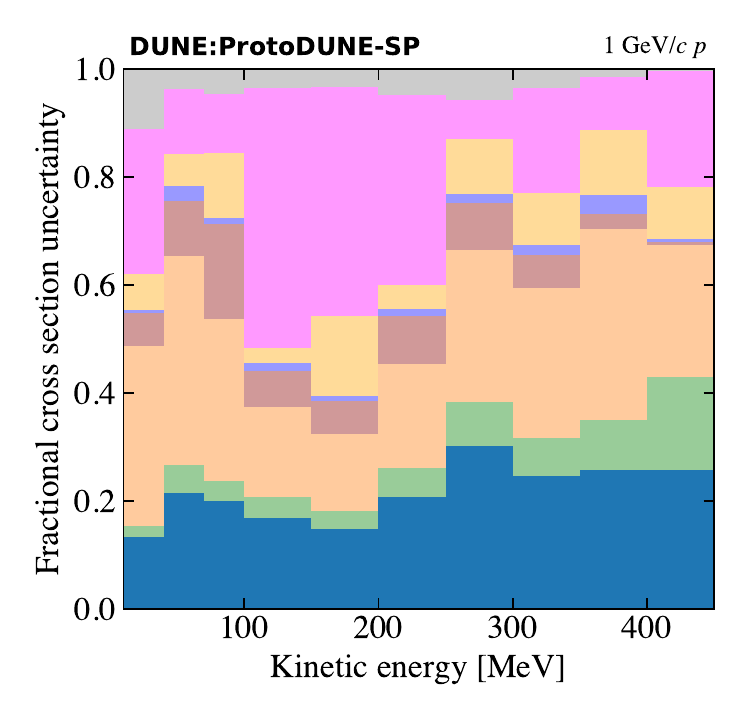}\put(-193,170){(b)}
    \caption{Different sources of cross section uncertainties across all energy bins for the (a) pion and (b) proton analyses.}
    \label{fig:uncertainties}
\end{figure*}

\begin{figure*}[htbp]
    \centering
    \includegraphics[width=0.45\textwidth]{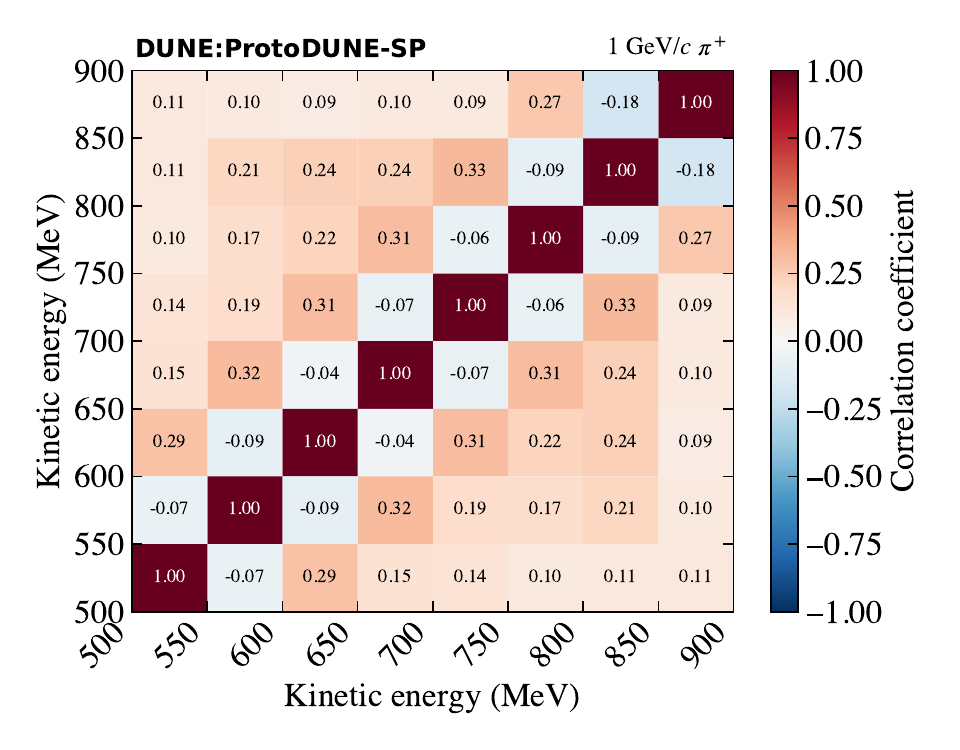}\put(-224,153){(a)}
    \includegraphics[width=0.45\textwidth]{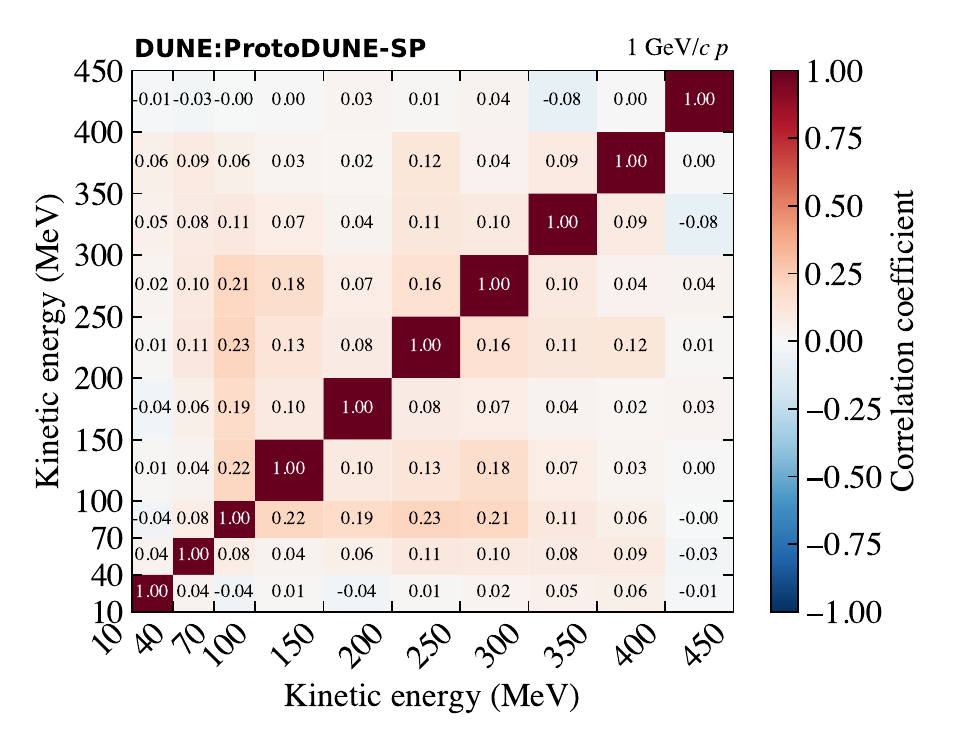}\put(-224,153){(b)}
    \caption{The correlation matrices for the measured results for the (a) pion and (b) proton analyses.}
    \label{fig:correlation}
\end{figure*}

The breakdown of total uncertainties into different sources across all energy bins is shown in Fig.~\ref{fig:uncertainties}. The correlation matrices for the measured results are shown in Fig.~\ref{fig:correlation}.

\end{document}